\documentclass{article}
\usepackage{graphicx}
\graphicspath{{figures/}}
\usepackage{appendix}
\usepackage{silence}
\usepackage{float}
\usepackage{subcaption} 
\usepackage[utf8]{inputenc}
\usepackage[T1]{fontenc}
\usepackage[english]{babel}
\usepackage{fullpage}
\usepackage{color}
\usepackage{url}
\usepackage{hyperref}
\usepackage[table]{xcolor}
\definecolor{darkWhite}{rgb}{0.94,0.94,0.94}
\usepackage{listings}
\begin{document}
	\title{Semantic Web Environments for MAS: Enabling agents to use Web of Things environments via semantic web\\[1cm]\small Internship report submitted for the   degree of \\master in computer science <<Cyber-Physical-Social Systems (CPS2)>> \\ Université Jean Monnet - Saint Étienne\\ August 2018. }
	%
	%
	\author{Alaa DAOUD\\[1cm]}
	\date{%
    Advisors:\\[.5cm]Maxime LEFRANCOIS , Antoine ZIMMERMANN  , Andrei CIORTEA, Olivier BOISSIER\\[1cm]Univ. Lyon, UJM, MINES Saint-Étienne, CNRS Lab Hubert Curien UMR 5516 F-42023 Saint-Étienne, France}
	%
	%
	%
	\maketitle              
	\begin{abstract}
		The Web is ubiquitous, increasingly populated with interconnected data, services, people, and objects. Semantic web technologies (SWT)
		promote uniformity of data formats, as well as modularization and reuse of specifications (e.g., ontologies), by allowing them to include and refer to information provided by other ontologies.
		In such a context, multi-agent system (MAS) technologies are the right abstraction for developing decentralized and open Web applications in which agents discover, reason and act on Web resources and cooperate with each other and with people.\\
		The aim of the project is to propose an approach to transform ``Agent and artifact (A\&A) meta-model'' into a Web-readable format with ontologies in line with semantic web formats and to reuse already existing ontologies in order to provide uniform access for agents to things.
		
	\end{abstract}
	\section{Introduction}
	The integration of devices into the Web facilitates the creation of a world-wide ecosystem of heterogeneous, loosely coupled things that lead to ease in mashing-up services using traditional web techniques to create hybrid WoT applications. While multi-agent systems facilitate the creation of goal-driven applications profiting from autonomy, collective intelligence, communication, organization, and other benefits that MAS technologies provide.
	\\
	Most present customer IoT devices are exposed to Web via cloud services developed by their manufacturers and many of them provide Web APIs. Developers may integrate functionality across heterogeneous things to create physical mashups that could be main building blocks in specific WoT application domains. But as this static mashups cannot simply adapt to dynamic IoT environments\cite{ciortea2017beyond}, We aim to simplify this process and bring it autonomy that supports discoverability of evolvable IoT environment,  in a way that the usage interfaces of IoT Things are accessed depending on their effects to environments and the description of desired goals, benefiting from semantic web technology to describe the Things' functionalities and also the desired goals.
	In Section \ref{usecases}, we present two scenarios to help illustrate the objectives of this project, then we explore some related work in AMAS \& Semantic Web communities in section  \ref{background}, in section \ref{contribution} we benefit from their work and approaches to provide our approach to decouple applications from the actual device interfaces they use. Then decouple software agents from these artifacts by providing a uniform description of their usage effects on environment, and in section \ref{discuss} we evaluate this approach by discussing different contexts  
	\section{Context and Motivation}
	\label{usecases}
	Several parallels have been made between research on multi-agent systems and the WoT domain, for example, in terms of interactions, communication protocols, interoperability and autonomous behaviors, where agents can interact with each other to find plans of using IoT Things to achieve goals on behalf of users.
We envision IoT objects as artifacts in multi-agent systems. Connected objects thus become "Web of Things-Artifacts".
\\Following scenarios illustrate our global vision considering Ted's Smart-Home, where Ted has equipped his house with smart devices to convert it to a smart home, thus each room was lightened with smart light bulbs, and equipped with a thermometer, air conditioner, fire detector, etc.
All these devices are connected to the local network that has Internet access, and accessible by their manufacturers APIs to be controlled via a smart-phone application.

In the beginning, everything went well. Ted was able to play some lighting scenarios, or choose his preferred protocols to handle event. For example when fire detected, by default a ringer alarm and SOS lighting in the fire location run, while he prefers to do emergency call when he is outside.\\
Such a system is simply realizable via traditional web mashups, now lets consider the following scenarios:
\subsection{Scenario 1: Device replacement}
\label{repscen}
Ted has an SOS lighting scenario that works correctly when he has a fire detected in some room, one day a light is broken. The functioning of the system adapts smoothly to this new configuration, redirecting the actions on this lamp to another one.
\\Ted replaces the broken light bulb with one from a different manufacturer, with a different API. Without any redesign of the smart app running on the smart phone, the system integrates this new light are adapts its functioning, recovering the previous quality of service.
\\This scenario addresses the need of discoverability when dealing with evolvable environment where devices appear, disappear, or change its response status dynamically at runtime, in addition to the need of dealing with huge heterogeneity of devices.
	\subsection{Scenario 2: Runtime configuration}
\label{confscen}
David uses similar system in his house, he has a protocol for welcoming home, where following this protocol will perform a set of actions to prepare a warm welcome in the next 30 minutes where he is expected to arrive, the system senses the temperature in hall and depending on this value and David’s preferred ambient temperature, his connected air conditioner (AC) start warming up or cooling, and romantic lighting scenario is applied.\\
David published this protocol script on some web service through the internet. Ted was interested in testing this welcoming home, but for air conditioning he has two kind of devices, one for air cooling and another for heating instead of one AC. He only passed that script’s URI to “import protocol” option on his smart phone application, thus he can load this protocol in his home upon his next arrival to home, then the system discovers and run all necessary functionalities of  devices to achieve this warm welcome.
\\In this scenario in addition to discoverability and heterogeneity handling, we have also the need for a platform that provides uniform access to things functionalities depending on their effects on environment.
\\\paragraph{}To achieve these scenarios in a stable system without a need for further hard coding via traditional web mashups, we need to model all possible IoT APIs for this application domain, provide stable environment discoverabilty modules that can handle all possible response statuses for these things, and to provide a huge database of functionalities effect on environment, these requirements are usually not feasible but if they were, they lead to inefficiency for small applications.
	\section{State of the art}
	\label{background}
	In this section, we’ll explore some research papers discussing the WoT application, MAS environments, and integration of semantic technologies in this field, and review this paper's objective, methodologies, results and their effect in the body of research in this field.\\ A starting point can be an overview of current approaches for building hybrid WoT applications, then we explore efforts from AAMAS community to abstract Agents environment that brought to an agent the possibility to act on real-world things in order to achieve autonomously user goals, and standardization efforts that aim to facilitate these approaches.

\subsection{WoT mashups and standardization efforts}
Most present customer IoT devices are exposed to Web via cloud services developed by their manufacturers and many of them provide Web APIs, Developers may integrate functionality across heterogeneous things to create physical mashups,\\ supposing we have already defined things' functionalities as APIs' endpoints, following a specific routine to compose more complex ones defines the process driven composition approach, where some researchers turned to the Web as an integration platform for things,[wilde 2007\cite{wilde2007putting}, Guinard 2009\cite{guinard2009towards}] proposed the first attempt to create physical mashups by developers that could be main building blocks in specific WoT application domains, while it is still required to build adapters for the APIs of each individual thing or WoT platform, [Rietzler et Al. 2013 «Homeblox»\cite{rietzler2013homeblox}] approaced the usage of mashup composition tools which don’t require programming skills, users can build their own mashups using simplified GUI with drag/drop and linking between icons via visual wiring of GUI widgets which is created manually by developers.
\\ Although developers are free to mashup different kind of services to implement new applications, and users are flexible to create there own scenarios, but no new mashup is created completely at runtime, because there is a static set of resources with static links that have been already defined by developers and users just instantiate these links, another limitation is that having a lot of functionalities, user GUI may become crowded at runtime which prevents users from creating new mashups thus we can say this kind of systems is relevant for scenarios with a limited set of resources, with no dynamic add/remove devices at runtime.
In other hand looking to all what is required to achieve the final goal, then decide the most relevant routine to do that, defines the goal-driven composition approach where we may follow a process, but we can break the current process and start over a new one. the main objective here is to reach the goal.
[Mayer 2014] proposed to use underlying infrastructure to deal with the automatic composition to achieve specific goals\cite{mayer2014configuration}, this approach implies using meta-model to describe services, while Users and possibly software clients specify their goals. this approach provides an effective way to implement such a goal-oriented composition of services when we have a limited set of resources, but when targeting dynamic, complex, and open environments, resources are added and removed at runtime, where Huge variety of heterogeneous Things have to be considered, standardization is needed in addition to autonomous decentralized decision making.
In 2018, Ciortea, Mayer and Michahelles\cite{ciortea2018repurposing} extended previous approach to design scalable and flexible agent-based production systems that integrate automated planning with multi-agent oriented programming for WoT, using semantic descriptions of Web-based artifacts  and coordinate through multi-agent organizations for manufacturing lines.
\\It is impossible to provide a complete overview of all the standardization efforts, some are collected here concerning IoT and WoT, we realize that this is only a small sample of relevant work.

\subsubsection{IOT.SCHEMA.ORG}
Schema.org community proposed to start working in IoT area, based on the existing schema.org success to provide a semantic interoperability approach that connects IoT and non-IoT applications. They describe some possible areas of application where relevant schema.org-oriented vocabularies can be useful, give some examples of relevant schemes and standard-related efforts, and suggest some practical steps for collaborative action.\cite{IOTSChemaweb}
\\Usually, schema.org "extension proposals" suggest a set of new types of things that can be added to the schema.org definitions, in addition to the typical characteristics of these types of objects. This proposal is quite different, it is only the beginning of the cooperation in these fields. Instead of suggesting a specific set of schema definitions, they suggest a broader set of collaborations whose results could include activities such as:

Develop appointments between the different application-oriented and platform-oriented schema and schema.org schema, through a shared data model.
Develop and document a common data model (such as RDF) capable of capturing various schemas.
\\
The publication of "External Extension" distribution systems based on the core of schema.org and common data model (for example, see http://gs1.org/voc/).
\\
Hosted Plugins - was published as part of the subdomains using the schema.org site, for example, see http://auto.schema.org/ or http://health-lifesci.schema.org/.
\\
Additions and modifications to basic schema.org terms (for example, adding a default to a base type, such as "device" or improvements on measures, procedures, and events).
\\
Collaborate on the improvement of examples, documents, sample codes, etc. related to Internet objects.\\\\
For the sake of simplicity, they call all these possibilities of cooperation on behalf of "IoT + schema.org" or "iot.schema.org". Due to the nature of the Internet of things, they expect the site iot.schema.org to be an entry point for a variety of activities, as well as to provide a home for any particular vocabulary to the IoT come out of cooperation.
\\\\
In the IoT environment, schema.org raises a number of considerations beyond simple modeling tasks. Some of the benefits of Internet applications are things that use the schema.org vocabulary for connected devices and services in that they rely on an existing task rule, so that they do not put an intolerable distinction between IoT/non-IoT scenarios/applications, and help decouple data structures from products and tools that use the data.
\subsubsection{W3C WoT Thing description}
To simplify application development and improve interoperability across different platforms, W3C recently (End of 2016) launched the Web of Things working group to develop Web standards to address the fragmentation of the Internet, reduce development costs, and reduce the risk for investors and customers. Encourage the rapid growth of the hardware and Internet services market.\cite{W3CWot}
\\The Internet facilitates the development of web applications regardless of technologies standing behind these things' implementation. W3C tries to do the same for the Internet of Things\cite{W3CWotIG}, To do so they addressed the need of a platform-independent API for developers uses the basis of information and interaction with rich descriptive metadata models of application and security requirements for communication platforms, that enables platforms to share the same meaning when they exchange data
They defined a Thing to be a representation of any physical or virtual entity must be represented in IoT applications. This entity can be a device, a logical component of a device, a local hardware component, or even an abstract virtual entity such as a location (for example, a room or a building). and thus they are working on standardizing
the initial building blocks to build the abstract structure of the WoT, and can be applied in three levels:\cite{ArchiW3CWot}
\begin{itemize}
	\item Device level;
	\item Gateway level (or "edge");
	\item Cloud level.
\end{itemize}

Things provide API for the WoT interface based on a formal model, objects can also be available on non-web protocols such as MQTT or ZigBee.\cite{W3CWot}
However, there may be things that do not provide a WoT interface and consist only of metadata related to the application (for example, the room where the hardware is located). In the WoT W3C, however, there must be a Thing Description (TD) that describes the thing. So, anything that has a TD is a thing.\cite{W3CWotTD}
\\ In W3C they are interested in 3 building blocks which are WoT thing description, WoT Binding Templates, and WoT scripting API. however, we are only interested now in WoT Thing Description (TD) which is defined as a data structure that adheres to a formal model and bridges the gap between Linked Data vocabularies and the functional APIs of IoT systems. It can be seen as the "HTML for things". A TD provides general metadata of a thing as well as metadata about the interactions, the data model, the communication and the security mechanisms of a thing. Typically, TDs use domain-specific metadata for which WoT provides explicit extension points. However, any domain-specific vocabulary is beyond the scope of the W3C standardization activity.\cite{W3CWotScript}

\paragraph{TD information Model}
The Thing Description information model is a set of basic definitions for the terms defined in [WOT-ARCHITECTURE]. These definitions take the form of a built-in vocabulary in the Resource Description Framework (RDF). An overview of this vocabulary with its class context and the class relation is given by the following three figures: the TD Core model, the TD data schema model, and the TD security model.
	\begin{figure}[ht]
     \centering
     \begin{subfigure}{\linewidth}
			\centering
			\includegraphics[width=0.7\linewidth]{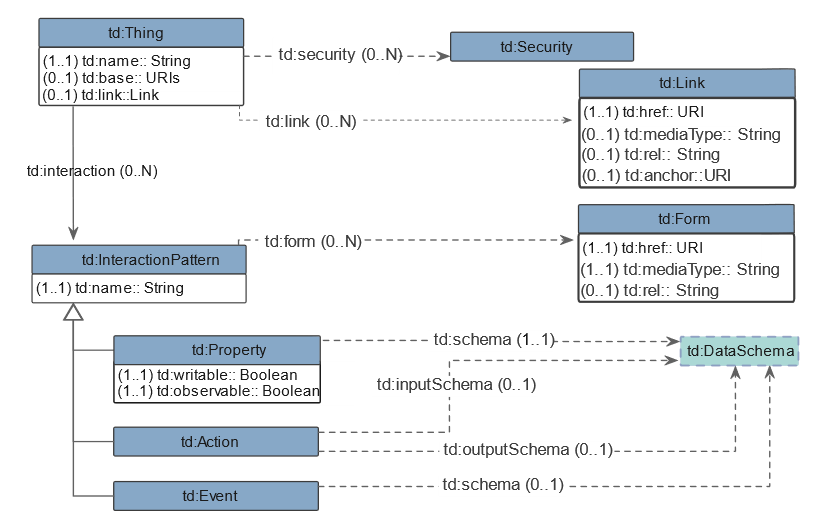}
			\caption{TD Core Model.}
         \label{fig:tdcore}
     \end{subfigure}
     \hfill
     \begin{subfigure}{0.4\linewidth}
			\includegraphics[width=\linewidth]{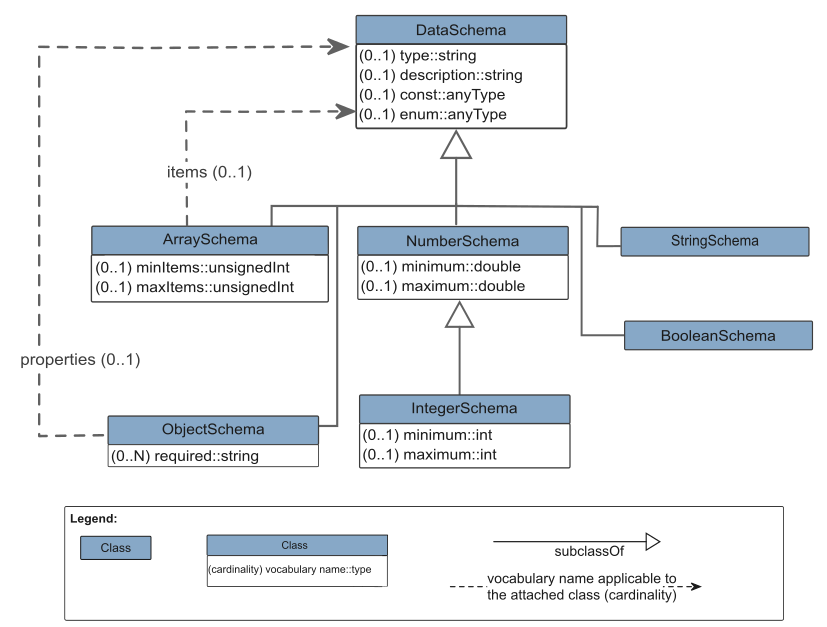}
			\caption{data schema model.}
         
         \label{fig:datasc}
     \end{subfigure}
     \hfill
     \begin{subfigure}{0.4\linewidth}
			\includegraphics[width=\linewidth]{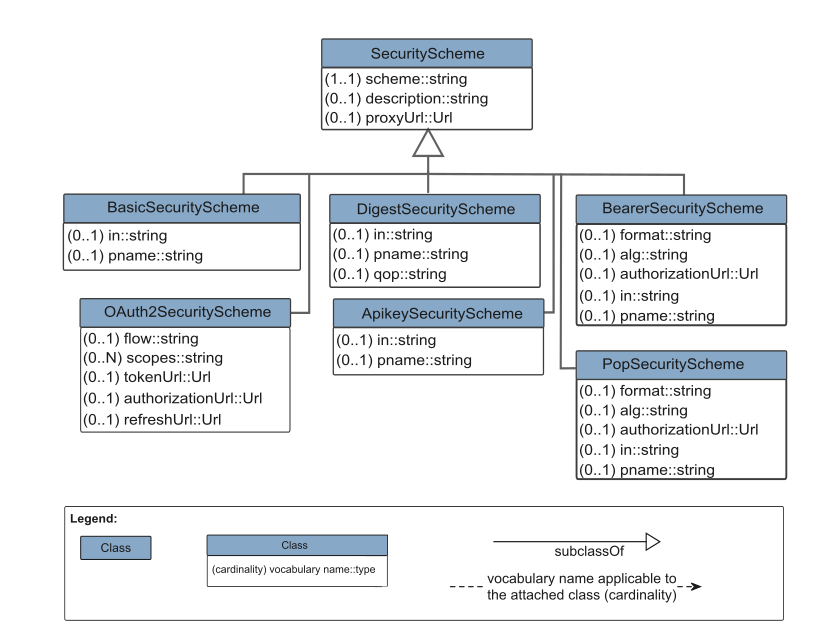}
			\caption{security model.}
         
         \label{fig:security}
     \end{subfigure}
        \caption{TD Information Model. W3C-WoT-TD\cite{W3CWotTD}}
        \label{fig:TDMODEL}
\end{figure}

\subsection{Environments for MAS}
Various models and architectures have been proposed for designing agent environments, and these designs have been validated in many applications. A special case is dedicated to agent environments in MAS in Autonomous Agents and Multi-Agent Systems Journal in 2007
\\
To define clear boundaries between the system and its environment in many cases, software systems need to interact with external software components, either to provide services or data or to achieve them.\cite{ricci2009environment}
\\
we can see a smart environment as a factor seen in the state of the environment using sensors and procedures on the environment using the controllers
\cite{russell2002artificial}

Since the software environment is very complex, there may be more devices, agents or systems that the environment serves.
Many groups have studied smart environment architecture for multi-agent systems.
When creating multi-agent perspectives, researchers should consider:
\begin{itemize}
	\item The role of the software which will be considered by each agent;
	\item The type of organization that exists between the agents;
	\item The method by which agent information will be shared.
\end{itemize}
Various software systems represented and modeled independently may "live" in the same environment and openly interact with others. Theosophical interactions require common ontologies, communication protocols, intermediate infrastructure and appropriate coordination to enable interoperability.

A series of workshops have been held from 2004, known as the (E4MAS 2004-2006), where the main objective was putting "Environments" on the Agenda of MAS research\cite{weyns2004environments}.\\
These workshop topics covered the entire scope of engineering environment from definitions to conceptual models, including a number of real-world applications.
The results of 3 successful workshops have been transferred to border community in different ways.

\subsubsection{Agent and Artifact (A\&A) Meta Model :}
\begin{figure}[ht]
	\includegraphics[width=\linewidth]{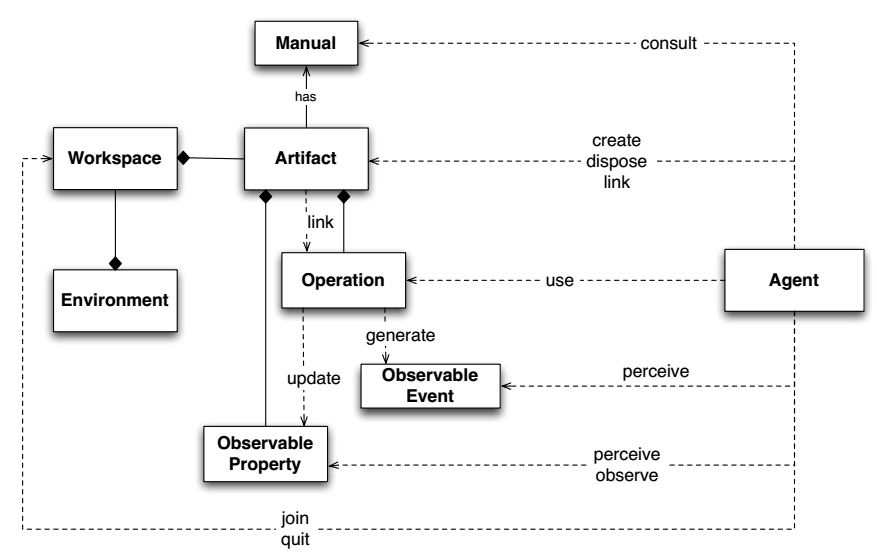}
	\caption{Agent and Artifact (A\&A) Meta model. (Ricci 2009\cite{ricci2009environment}) }
	\label{fig:cartago}
\end{figure}

In 2009 Ricci\cite{ricci2009environment} proposed Agent and Artifact (A\&A) Meta Model (see figure \ref{fig:cartago}) that makes clear separation between non-autonomous entities (Artifacts) which are dynamically constructed resources, support individual \& collective activities, and proactive entities (Agents) which are goal oriented entities, can discover, sense and use artifacts to achieve their design objectives, thus in this context environment is designed as a dynamic set of artifacts

	\section{Agent, Artifact and Thing (AAT) Model}
	\label{contribution}
	We aim to bring autonomy to WoT Applications, considering openness and evolvability of the system, where must be taken into account the huge variety of heterogeneous platforms and APIs for IoT Things coming from competitive manufacturers, that make it difficult to share platform independent meanings between these things, to provide hybrid WoT application runs effectively in an open, evolvable and dynamic environment, and able to be auto-configured, self adapted when it discovers any change in the dynamic context of its environment, including the availability status of things.
	\begin{figure}
		\centering
		\includegraphics[width=0.7\linewidth]{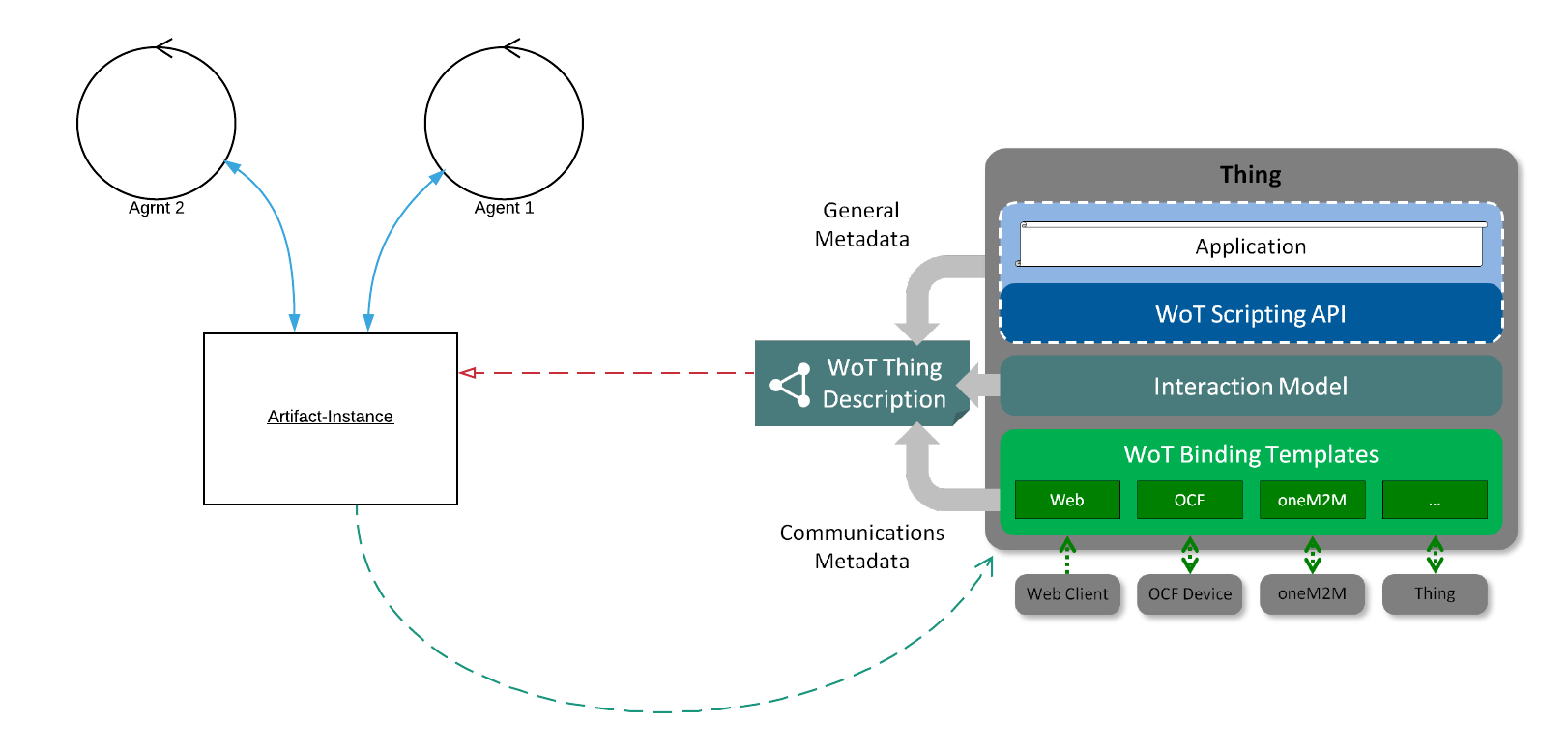}
		\caption{Agent,Artifact\&Thing (AAT) abstract model}
		\label{fig:objective}
	\end{figure}
	
	To provide such a solution we suggest an approach to extend Mayer's approach\cite{mayer2014configuration} on two levels:
\begin{itemize}
	\item The ability to configur on the fly: Benefiting from standard description of IoT things in semantic web manner, to decouple the representation of these things in the system (by means of artifacts) from their real-world implementation
	\item Autonomy in decision making: To achieve defined goals, using a semantic description of artifact usage interface (by means of a semantic manual) to decouple artifacts from its usage interface that is accessed by agents
\end{itemize}
Thus we can achieve a generic purpose platform for developing IoT application benefiting from standardization efforts done by W3C and iot.schema.org, and the facilities that MAS provide especially A\&A based platforms to deal with IoT things.
\subsection{Decoupling artifacts from real-world devices}
\begin{figure}[ht]
	\centering
	\includegraphics[width=\linewidth]{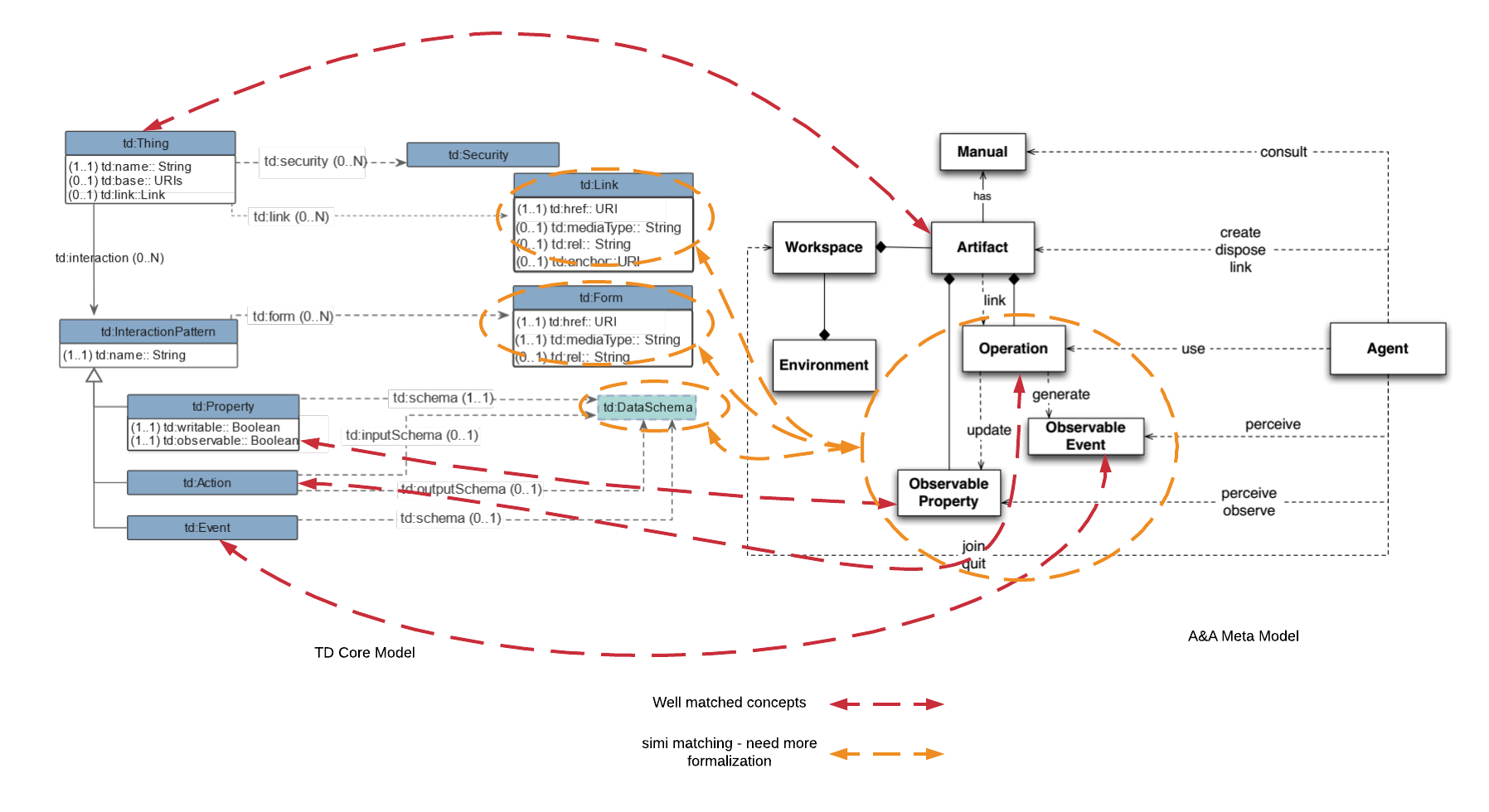}
	\caption{conceptual similarity between TD Core Model \& A\&A Meta Model}
	\label{fig:mapping}
\end{figure}

The core model of W3C's WoT Thing Description\cite{W3CWotTD} and Artifact in  Ricci's (A\&A) meta model\cite{ricci2009environment} were proposed for very different purposes, although both conceptually present environment entities attributes, relations, and interaction models by means of (Properties, Events and Actions/Operations). That leads to unignorable  conceptual similarities between these models, figure \ref{fig:mapping} shows the conceptual mapping between these models. 

\begin{figure}[ht]
	\centering
	\includegraphics[width=0.4\linewidth]{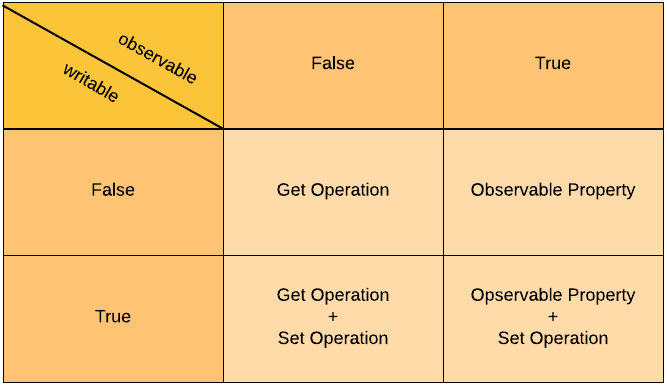}
	\caption{td:Property constraints}
	\label{fig:prop}
\end{figure}

Benefiting from this similarity, we propose generic artifact type with parametric polymorphism usage interface. This artifact can be instantiated from any TD of a real-world device (Thing) to an artifact instance that represents this thing in agent understandable manner, enabling:
\begin{itemize}
	\item Dynamic definition of observable properties
	\item Dynamic set of operations
	\item Dynamic set of observable events
\end{itemize}
This addresses the need of gathering information from TD file and provide it to the generic artifact to be instantiated.
\\
Because of the heterogeneity of IoT Things manufacturers, there is no agreement on a standard communication protocol to call Thing's functionalities, thus we need a protocol binding handler that deals with different protocols (eg HTTP, CoAP\cite{shelby2014constrained}, MQTT, etc.) to access interaction patterns of Things, construct the right request payload, and send it via the right protocol method then handle response. Where information about that communication and data schema for payload is provided by TD in the object of td:form predicate.
\begin{figure}[ht]
	\centering
	\includegraphics[width=0.7\linewidth]{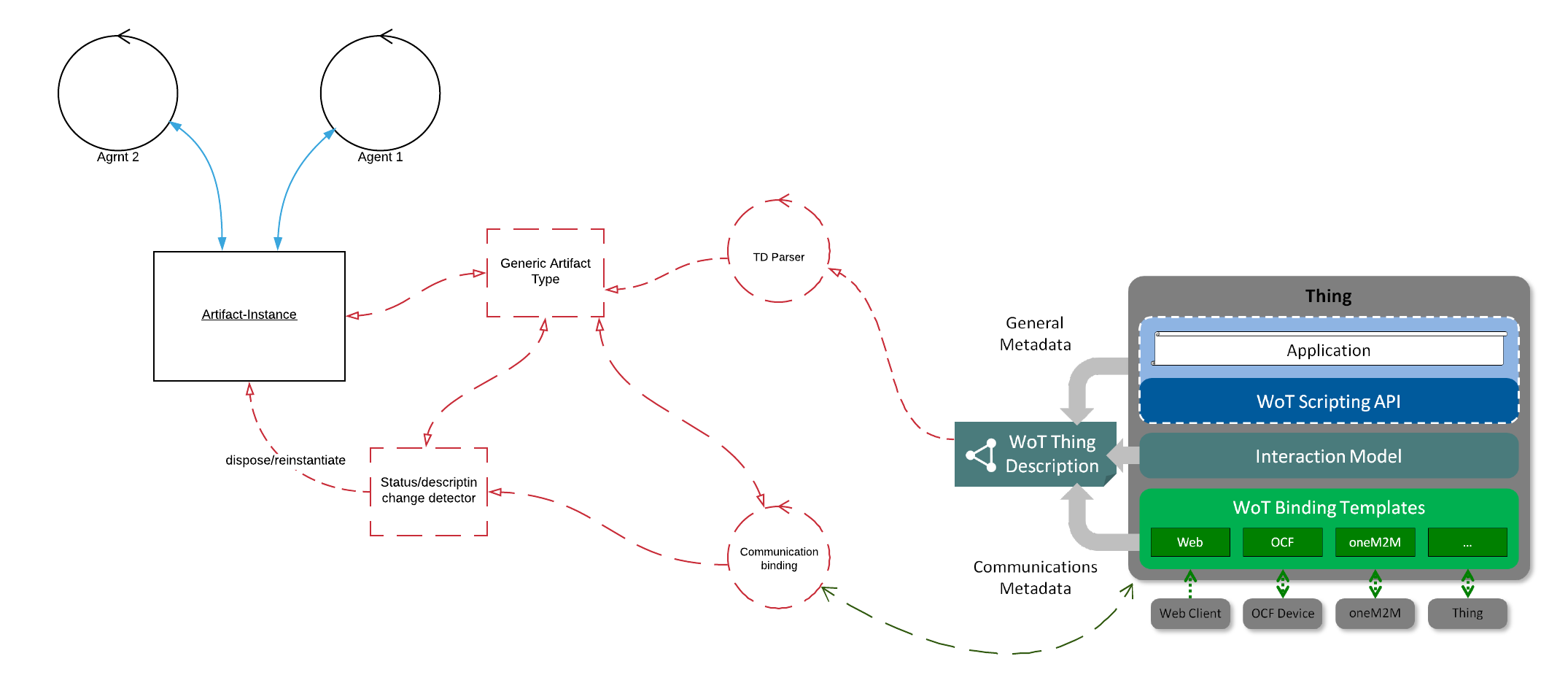}
	\caption{Generic artifact model}
	\label{fig:aat}
\end{figure}
Figure \ref{fig:aat} shows this generic artifact model, where agents interact with the real world devices via instances created from generic artifact type by parsing its TD, and communication passes through its defined protocols via communication binding handler.
\\Let's say we have an IoT device ``dev1'' described in ``TD1'' where it has ``td:name=dev1name'', the parser will read ``TD1'' to instantiate generic artifact to have an instance named ``dev1nameArtifact'' that provides access through Operations to all dev1 interaction patterns of type td:Action defined in TD1, for those of type td:Property it create observable\_property or object field with getter and/or setter regarding writable \& observable constraints of this pattern (see figure \ref{fig:prop}) \\
In this model the user can at runtime add new device by providing its proper TD, and replace an existing device with a new one, no matter if it is from the same manufacturer or not, no matter if it uses the same protocols or not, while this information is provided properly in the new TD, the protocol binding and payload handler deals with constructing the right requests to its API.
\\ Note that agents must have some beliefs about artifacts and their operations to look for alternatives to execute their plans. Suppose we have an agent plan to trigger the operation ``Op1'' of the artifact ``art1''. At some point the user replaced the device which is represented in ``art1'' with another device, he is obliged to give the new device the same (name/base IRI) as previous one, and give the new operation the same (names/IRI patterns) of there equivalent in the previous TD, to stay able to use the same agent plan. Otherwise agent plans should be modified to match the new environment. In the following section, we propose a solution to this problem.

\subsection{Decoupling agents from artifacts}
\label{usageLogic}
In open evolvable IoT environments, Things may appear, disappear, or be replaced dynamically. And Regarding the heterogeneity of Things manufacturers there is no standards for device or operation naming, that leads to expire agents beliefs regarding these things at runtime (as described in the previous section).
\\We aim to deal with artifact operations in terms of their effects on environment context, thus agent plans will base on discovering possible operations to call, regarding the matching between their specified preconditions and the current context of environment, and chose the ones that lead to the desired goal regarding their postconditions as described in the following logic:
	
	\begin{itemize}
		\item \(Operation(O)\) : O is an operation 
		\item \(Artifact(A)\) : A is an artifact 
		\item \(Usage(U) \) : U is a usage protocol
		\item \(Context(C)\) : C is an environment context
		\item \(Status(S)\) : S is a status statement
		\item \(Predicate(P)\) : P is a predicate type
		\item \(Goal(A)\) : A defines desired context
		\item \(instancOf(I,T)\) : Item \(I\) is an instance of type \(T\)
		\item \(hasProperty(C,S)\) : status  \(S\) is on of the properties of context \(C\) 
		\item \(precond(U,C)\) : the usage \(U\) has the context \(C\) as precondition
		\item \(postcond(U,C)\) : the usage \(U\) has the context \(C\) as postcondition
		\item \(forArtifact(U,A)\) : uage  \(U\) is only valid for artifacts of type \(A\)
		\item \(hasOperation(A,O)\) : artifact of type \(A\) has an operation of type \(O\)
		\item \(forOperation(U,O)\) : usage \(U\) is valid for operations of type \(O\)
		\item \(propertyOF(S,A)\) : status \(S\) describes a property of artifact of type \(A\)
		\item \(propertyOfInstance(S,I,A)\) : status \(S\) describes a property of artifact instance \(I\) of type \(A\)
		\item \(hasValue(S,P,V)\) : Status statement \(S\) specify a predicate of type \(P\) that has value \(V\)
		\item \(self(S)\): Status statement S is restricted to property of artifact instance not to artifact type in general
	\end{itemize}
	In this context we can define following statements to be always valid for any instance of this model:
	\[ \forall C : Context(C)\Rightarrow \exists S: Status(S)\wedge hasProperty(C,S)\]
	\[ \exists A\exists O: Artifact(A)\wedge Operation(O)\wedge hasOperation(A,O)\]
	\[ \exists U \exists C: Usage(C)\wedge Context(C)\wedge precond(U,C)\]
	\[ \forall U : Usage(U)\Rightarrow \exists C: Context(C)\wedge postcond(U,C)\]
	\[ \forall U : Usage(U)\Rightarrow \exists A: Artifact(A)\wedge forArtifact(U,A)\]
	\[ \forall U : Usage(U)\Rightarrow \exists O: Operation(O)\wedge forOperation(U,O)\]
	\[ \forall S : Status(S) \Rightarrow \exists P\exists V : Predicate(P)\wedge hasValue(S,P,V)\]\
	\[\exists S: Status(S) \wedge self(S)\]
	\[\forall G : Goal(G) \Rightarrow Context(G)\]
	To determine if a goal context can be achieved via a usage of some instance of artifact (regardless preconditions)\\\\
	\(\forall G\forall U\forall I: hasProperty(G,S1)\wedge postcond(U,PC)\wedge hasProperty(PC,S2)\wedge instanceOf(I,A)\wedge propertyOF(S1,A)\wedge propertyOF(S2,A)\wedge\neg self(S2)\wedge hasValue(S1,P,V)\wedge hasValue(S2,P,V)\Rightarrow acheivable(G,U,I)\)
	\\\\Here the goal is achievable for any instance of artifact type that satisfy previous constraint, while:
	\\\\
	\( \forall G\forall U : hasProperty(G,S1)\wedge postcond(U,PC)\wedge hasProperty(PC,S2)\wedge propertyOF(S2,A)\wedge self(S2) \Rightarrow\)
	\\
	\( (\forall I: propertyOfInstance(S1,I,A)\wedge hasValue(S1,P,V)\wedge hasValue(S2,P,V) \Rightarrow acheivable(G,U,I))\)
	\\\\
	determines that there is only some instances that can achieve the goal when it is restricted to its own properties.
	\\ More complex goals can be defined as a union of desired contexts, that can be achieved by achieving all its components
	
\section{Implementation}
The implementation presented here should be considered proof of the concept and does not use all the aspects presented in section \ref{contribution}. However, we consider it a good starting point to explain the use of our approach. Our goal is to show that the dynamic configuration of the infrastructure can be managed at runtime without modifying agent or artifact codes.
\\We use the Java programming language for this implementation using JaCaMo\cite{boissier2013multi} framework for MAS programming that combines three separate techniques, Jason as an interpreter for extended version of AgentSpeak for the development of multi-agent systems, CArtAgO as a general purpose framework that makes it possible to program and execute virtual environments for multi-agent systems based on Agent \& Artifact (A\&A) metamodel, and Moise for programming multi-agent organisations,
Ontology was built using Protegé, and for parser we used apache Jena that enables creating in-memory RDF model from RDF file written in any format (JSON\_LD in our case) then gathering information via SPARQL queries on that model, which make this parser format independent where the most important is the information triples contained in the model and namespaces that are used to bind concepts URIs.
\subsection{Generic Artifact}
\begin{figure}
	\centering
	\includegraphics[width=0.5\linewidth]{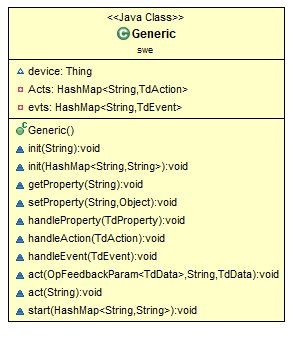}
	\caption{Generic artifact type(class)}
	\label{fig:generic-artifact}
\end{figure}

The generic artifact type (shown in figure \ref{fig:generic-artifact}) which is written in CArtAgO, can be instantiated via URI of a TD, the init operation uses the parser to get an object structure named device that represents this thing (see figure \ref{fig:thingcd} in Appendix \ref{classes}), based on this structure it determines what are the Observable\_Properties, Operations, and Actions that will be included in the instance, it adds all Observable properties dynamically via defineObsProperty(name,value), puts all Operations/Actions in Acts HashMap indexed by operation name, Events in evts HashMap in the same way.
\\This Java-Class/Artifact-Type provides a uniform usage interface to trigger Actions via a uniform operation act(OpFeedbackParam,actionName,inputData), where actionName is a String that is a key of Acts map, OpFeedBackParam and inputData are objects of type TdData (see figure \ref{fig:data} in Appendix \ref{classes}) that follows input/output schema identified in the action form.
Returning back to device replacement scenario, Teds SOS lighting scenario uses emergency\_light artifact which represent light bulb LB1 provided by manufacturer A and described in listing \ref{lst:httpTD}, this artifact is of type iot:Light and has (switch\_on/switch\_of) operations, thus the agent plan that handles this scenario is simply periodical call of these two operations sequentially as seen in listing \ref{sosPlan}.
\\Ted replaces the broken LB1 with LB2 from Manufacturer B, no matter if B = A or not. While LB2 is of type iot:Light and has (switch\_on/switch\_of) operations even called via different APIs or different protocols.\\Without any redesign of the smart app running on the smartphone, he just loads the TD of LB2, giving it the same td:name emergency\_light in listing \ref{coapTD}.\\ The system re-instantiate in runtime the emergency\_light artifact with the new TD, thus agents can still use it as before but for the new bulb, the usage interface doesn't change while protocol binding and payload handler modules deals with constructing the right requests to the new API.

\begin{lstlisting}[label=lst:httpTD, caption=old lamp TD\\]
{
"@context": [
	"https://w3c.github.io/wot/w3c-wot-td-context.jsonld",
	"https://w3c.github.io/wot/w3c-wot-common-context.jsonld",
	{
		"iot": "http://iotschema.org/",
		"http": "http://iotschema.org/protocol/http"
	}
	],
"@type": [ "td:Thing", "iot:Light"],
"td:base": "http://localhost/TD",
"td:name": "emergency_light",
"interaction": [
	{
	"td:name": "Switch State",
	"@type": ["td:Property", "iot:SwitchStatus"],
	"td:schema":  {"@type": ["iot:SwitchData"], "type": "boolean"},
	"td:form": [{"href": "/currentswitch", "rel": ["readtd:Property"], "mediaType": "application/json"}]
	},
	{
	"td:name": "Switch On",
	"@type": ["td:Action", "iot:SwitchOn"],
	"inputSchema":  {"type": "boolean"},
	"td:form": 
	[{ "href": "/switchOn", "rel": ["invoketd:Action"], "mediaType": "application/json"}]
	},
	{
	"td:name": "Switch Off",
	"@type": ["td:Action", "iot:SwitchOff"],
	"inputSchema":  { "type": "boolean"},
	"td:form": 
	[{"href": "/switchOff", "rel": ["invoketd:Action"], "mediaType": "application/json"}]
	}
	]
}\end{lstlisting}
\begin{lstlisting}[label=sosPlan, caption=agent plan for sos light]
+soslight : fireEvent
	<- while(true) { 
		.startOperation("Swich On")[artifact_name("emergency_light")];
		.wait(2); 
		.startOperation("Swich Off")[artifact_name("emergency_light")];
		.wait(2); 
	}\end{lstlisting}
\begin{lstlisting}[label=coapTD, caption=new lamp TD\\]
{
"@context": [
	"https://w3c.github.io/wot/w3c-wot-td-context.jsonld",
	"https://w3c.github.io/wot/w3c-wot-common-context.jsonld",
	{
		"iot": "http://iotschema.org/",
		"coap": "http://iotschema.org/protocol/coap"
	}
],
"@type": [ "td:Thing", "iot:Light"],
"td:base": "coap://exampleHost/light",
"td:name": "emergency_light",
"interaction": [
	{
	"td:name": "Switch State",
	"@type": ["td:Property", "iot:SwitchStatus"],
	"td:schema":  {"@type": ["iot:SwitchData"], "type": "boolean"},
	"td:form": [{"href": "/currentswitch", "rel": ["readtd:Property"], "mediaType": "application/json"}]
	},
	{
	"td:name": "Switch On",
	"@type": ["td:Action", "iot:SwitchOn"],
	"inputSchema":  {"type": "boolean"},
	"td:form": 
	[{ "href": "/switchOn", "rel": ["invoketd:Action"], "mediaType": "application/json"}]
	},
	{
	"td:name": "Switch Off",
	"@type": ["td:Action", "iot:SwitchOff"],
	"inputSchema":  { "type": "boolean"},
	"td:form": 
	[{"href": "/switchOff", "rel": ["invoketd:Action"], "mediaType": "application/json"}]
	}
  ]
}\end{lstlisting}
These results prove that this approach can work if agents have beliefs about artifacts and their operations, it requires the agent to know that there is an artifact that has two operations ``Switch On''\ ``Switch Off'', it was required to give the new artifact the same td:name or the same Base IRI used for the old one or have an access to some resource directory that can be used to find an artifact of type "iot:Light" located in some specific location. If the agent couldn't find the artifact or the artifact operation names changed, agent can't run this plan.
\subsection{Usage Ontology}
To realize the usage logic that we proposed in section \ref{usageLogic} we developed an OWL Ontology to describe usage interfaces in terms of their pre/post-conditions, artifact they belong to, and artifact operation they describe. (see Appendix \ref{onto} for more details)
\begin{figure}[ht]
	\includegraphics[width=\linewidth]{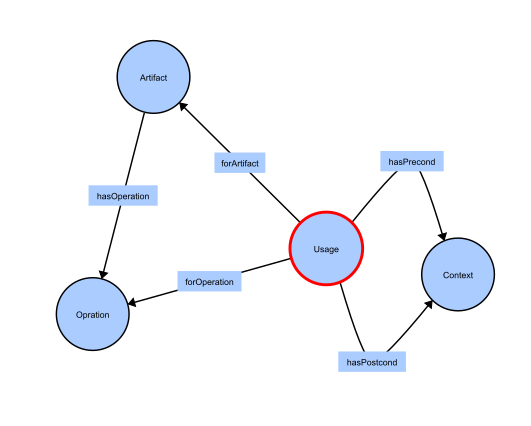}
	\caption{Usage Ontology}
	\label{fig:ontology}
\end{figure}
In this context the agent plan to run SOS lighting scenario should try to find two operations of the same light artifact, that could be called sequentially to switch this artifact status on/off, i.e. look up two usages U1 \& U2 both for the same artifact A1, which is an instance of iot:Light, and has two operation Op1, Op2, where U1 is forOperation Op1 \& U2 is forOperation Op2 \& U1 hasPostcond \{A1 iot:switch\_status "On"\} \& U2 hasPostcond \{A1 iot:switch\_status "Off"\}.
\\Thus no matter if a light bulb has broken or became unavailable at runtime, this plan can achieve the desired goal if it successes to find operations with the same effects on the environment.
\\
More complex IoT scenario can be defined as a set (or sequence) of IRIs, each refers to a Context RDF graph (Goal) to be achieved, so agent plan/s, in this case, try to lookup usages that achieve these goals one by one.
\\The runtime configuration scenario is achievable by this approach, David has protocol script that describes his welcoming-home as a sequence of IRIs each refers to a goal context defined somewhere on the web, he published his script through a web service, thus it is now accessible via its IRI, all that Ted needs to try this protocol in his home is to load this script, so that he has on the fly defined a sequence of goals as his desired scenario that he can set it as a response to his arrival or any other preferred event.

	\section{Discussion}
	\label{discuss}
	In this section we evaluate the approach with several examples in different levels of complexity, in order to find its weak points, and to address where this approach has to be extended or enriched, we start the evaluation with simple settings to show first its main advantages, then evaluate more complex settings in different levels 
	\subsection{Simple Settings}
	An example of a simple usage description :
	\begin{lstlisting}
	@prefix iot: <http://iotschema.org/>.
	@prefix rdf: <http://www.w3.org/1999/02/22-rdf-syntax-ns#>.
	@prefix usg: <http://www.emse.fr/ci/ontologies/2018/wot_usage#>.
	@prefix tools: <http://localhost/tools/>.
	
	_:switchOnUsage		a			usg:Usage;
				usg:hasPrecond		tools:lightOffContext;
				usg:hasPostcond		tools:lightOnContext;
				usg:forArtifact		_:lightArtifact;
				usg:forOperation	_:switchOnOperation.
				
	_:lightArtifact		a			iot:Light;
				usg:hasOperation	_:switchOnOperation.
				
	_:switchOnOperation	a		iot:SwitchOn.
	\end{lstlisting}
	where the context that describes the preconditions of this usage defined by "lightOffContext":\\
	\begin{lstlisting}
	_:lightArtifact iot:switchstatus	"off";		
	\end{lstlisting}
	
	while the context that describes the postconditions of this usage defined by "lightOnContext":\\
	\begin{lstlisting}[language=C]
	_:lightArtifact iot:switchstatus	"on";	
	\end{lstlisting}

	To define smart home structure we used Building Topological Ontology ``http://www.w3id.org/bot" as follows where zones define building parts and the elements that are contained by these zones are defined by URIs of their TDs:
	\begin{lstlisting}
@prefix iot: <http://iotschema.org/>.
@prefix rdf: <http://www.w3.org/1999/02/22-rdf-syntax-ns#>.
@prefix usg: <http://www.emse.fr/ci/ontologies/2018/wot_usage#>.
@prefix bot: <http://www.w3id.org/bot#>.
@prefix	sh:	 <http://localhost/smart_home#>.

sh:my_home	a	bot:Zone;
	bot:containsZone	sh:my_kitchen, sh:room1, sh:room2, sh:hall.

sh:my_kitchen 	a	bot:Zone;
	bot:hasElement	<http://localhost/TD/smart_home/kitchen/ceilingLight.jsonld>.
	\end{lstlisting}
	so we can define the context of the environment at any point of time in an arbitrary way as follows
	\begin{lstlisting}
<http://localhost/TD/smart_home/kitchen/ceilingLight.jsonld> iot:switchstatus "off".
<http://localhost/TD/smart_home/kitchen/emergencyLight.jsonld> iot:switchstatus "off".
<http://localhost/TD/smart_home/kitchen/curtains.jsonld> iot:currentStatus "closed".
<http://localhost/TD/smart_home/dining_room/d_ceilingLight.jsonld> iot:switchstatus "on".
<http://localhost/TD/smart_home/dining_room/decorationLight.jsonld> iot:switchstatus "off".
<http://localhost/TD/smart_home/dining_room/curtains.jsonld> iot:currentStatus "closed".
sh:my_kitchen	_:light_outside		true;
				_:brightness		"low".
	\end{lstlisting}
	and the goal we want to achieve :
	\begin{lstlisting}
<http://localhost/TD/smart_home/kitchen/ceilingLight.jsonld> iot:switchstatus  "on".
	\end{lstlisting}
	In this very simple example we can see that
	\[lightOffContext \models Context\cup Topology\]
	\[lightOnContext \models goal\cup Topology\]
	Thus in terms of preconditions the usage(SwitchOnUsage) of (switchOnOperation) for LightArtifact(CielingLight) is valid in the current Context.
	\\ In terms of postconditions, this usage can achieve the desired goal, so a valid solution to achieve the desired goal from the current context is provided by SwitchOnUsage.
	\\
	Assuming we have 2 light artifacts in the kitchen as follows
	\begin{lstlisting}
sh:my_kitchen 	a	bot:Zone;
	bot:hasElement	<http://localhost/TD/smart_home/kitchen/ceilingLight.jsonld>,
					<http://localhost/TD/smart_home/kitchen/emergencyLight.jsonld>.
	\end{lstlisting}
	The simple entailment method will lead to usages of SwitchOnOperation for both artifacts as solutions to achieve the desired goal. This is correct in terms of semantics because there is no information in those models refers to which artifact the effect will take place, but that can't help us in terms of selecting the right plan to achieve our goal.\\
	We suggest to use the predicate tools:referencedBy to define a reference for blanc nodes, thus we can know at some point that ``\_:lightArtifact" in both pre/postCondition models refers to the same thing, i.e. to use the switchOnOperation of ceiling\_light we have to check if the status of ceiling\_light is "off" as precondition and we then know that this status will change to "on" after this operation.
	\begin{lstlisting}
		_:lightArtifact iot:switchstatus	"off";
						tools:referencedBy tools:lightArtifact1.
	\end{lstlisting}
	\begin{lstlisting}
		_:lightArtifact iot:switchstatus	"on";
						tools:referencedBy tools:lightArtifact1.
	\end{lstlisting}
	and also the usage model will look like:
	\begin{lstlisting}
_:switchOnUsage		a			usg:Usage;
		usg:hasPrecond		tools:lightOffContext;
		usg:hasPostcond		tools:lightOnContext;
		usg:forArtifact		_:lightArtifact;
		usg:forOperation	_:switchOnOperation.
		
_:lightArtifact		a		iot:Light;
		usg:hasOperation	_:switchOnOperation;
		tools:referencedBy	tools:lightArtifact1.
		
_:switchOnOperation	a		iot:SwitchOn.
	\end{lstlisting}
	Thus we know that the switchStatus of the artifact instance that we trigger its operation is the one that we have to check for preconditions, and to be changed in terms of postconditions, and instead of doing full union we select only the relevant triples from usage/Context/Topology, etc. thus prevent ambiguity in graph matching and entailment.
	\subsection{More Complex Settings}
	The previous section viewed the main advantage of using this approach to have a semantic manual for artifact operation usages, but usually the goals are more complex than the provided example, and also the context, preconditions, and postconditions are complex, thus simple entailment cannot be achieved in most cases. for example when we have this goal:
	\begin{lstlisting}
<http://localhost/TD/smart_home/kitchen/light1.jsonld> 	iot:switchstatus	"off".
<http://localhost/TD/smart_home/kitchen/curtains.jsonld> iot:status		"open".
	\end{lstlisting}
	there is an operation usage (SwitchOnUsage) that has directly postcondition that implies:
			\[lightOnContext \models goal\cup Topology\]
	and thus we know that we can call this operation, but will the goal be achieved after this call? \\
	However, we still have a part of the goal that isn't achieved which is  
	\begin{lstlisting}
	<http://localhost/TD/smart_home/kitchen/curtains.jsonld> iot:status		"open".	\end{lstlisting}
	
	Another example of complexity when we have postcondition graph of a usage (SwitchOffUsage) like:
	\begin{lstlisting}
		_:kitchen		iot:brightness		"low".
		_:lightArtifact iot:switchstatus	"off".	\end{lstlisting}
	And the goal is defined as
	\begin{lstlisting}
	<http://localhost/TD/smart_home/kitchen/light1.jsonld> iot:switchstatus	"off".	\end{lstlisting}
	We know that the goal can be achieved by SwitchOffUsage, but the entailment method won't suggest such a solution because 
	\[postcondition \not\models goal\cup Topology\] as it has side effect triples.\\
	In many cases both previous situations happen at the same time, complex goal and side effects of usage:
	\\Goal:\begin{lstlisting}
<http://localhost/TD/smart_home/kitchen/light1.jsonld> 	iot:switchstatus	"off".
<http://localhost/TD/smart_home/kitchen/curtains.jsonld> iot:status		"open".	\end{lstlisting}
	postconds:\begin{lstlisting}
_:kitchen		iot:brightness		"low".
_:lightArtifact iot:switchstatus	"off".	\end{lstlisting}
In such settings, it is almost impossible to find direct solution. We suggest to follow the problem-solving heuristic, with the intent of finding a specific sequence of stages, that at each stage an operation usage is applied, and at the end we reach a final context that includes totally the goal context.
In each stage we determine all possible operation usages where their preconditions matches the context (current state of the environment), then predict the provisional context of the next step by using update SPARQL queries on context model.
\\The SPARQL 1.1 update provides these graph update operations\cite{SparqlW3C}:
\begin{itemize}
	\item The INSERT DATA operation adds triples, given online in the query, in a graph. This SHOULD create the destination graph if it does not exist. If the graph does not exist and it can not be created for any reason, a failure MUST be returned.
	\item The DELETE DATA operation deletes some triplets, given online in the query if the respective graph contains them.
	\item The fundamental pattern-based actions for graph updates are INSERT and DELETE (which can coexist in a single DELETE / INSERT operation). These actions consist of groups of triplets to delete and groups of triples to add. The specification of triplets is based on query templates. The difference between INSERT / DELETE and INSERT DATA / DELETE DATA is that INSERT DATA and DELETE DATA do not replace bindings in a model with a template. DATA forms require concrete data (triple models containing variables in DELETE DATA and INSERT DATA operations are not allowed and blanc nodes are forbidden in DELETE DATA).
	\item The LOAD operation reads the contents of a graph document in a chart in the graph store.
	\item The CLEAR operation removes all triplets in (one or more) graphs.
\end{itemize}
In our case, the INSERT clause should contain all triples in the postcondition graph after replacement of blank nodes with real IRIs from the current stage context and/or variables, while the preconditions graph triples are used to construct DELETE and WHERE clauses with the same binding of variables and/or IRIs.
\\The resulting graph of this operation represents the provisional stage context that is used as current context to calculate the next stage.
\\If some stage reaches a provisional graph that includes the goal graph, we know that the sequence of usages led to this stage can be used to achieve the goal.\\

In such settings, the problem is converted into a Path-Finding-like problem, several algorithms and techniques are able to solve this kind of problems, and choosing the best one is a research topic itself depending on the implementation objectives, thus we won't go in details, but we'll mention in what follows some of these algorithms

\begin{itemize}
	\item Ignore side effects and get the union of all usages that participate in achieving the desired goal
	\item Convert to an optimization problem, maximize achievable part of the goal in each step (Dijkstra, Greedy, Integer linear programming, DCOP, ... etc.)
	\item Divide and conquer algorithms: divide goal context into subgoals till very simple triples thus they can be achieved via the first simple approach
\end{itemize}
	\section{Conclusion}
	As the integration of Things into the Web facilitates the creation of hybrid WoT applications, We addressed the limitation of traditional techniques for physical mashups, then proposed our approach to provide a uniform usage interface for IoT things benefiting from W3C-WoT Thing Description, and autonomy of MAS especially  
	\newpage
	\bibliographystyle{plain}
	\bibliography{ref}
		\newpage
	\appendix
	\appendixpage
	
	\section{Definitions}
	we present the most common definitions of used technologies in this introduction, then some usecases that illustrate the aim of this project
	\paragraph{Multi-Agent-Systems (MAS)} is defined as a computerized system composed of multiple agents, situated in shared environment, interact with one another to achieve their objectives, suitable for building open, distributed systems situated in dynamic and complex environments, and may interact or act on behalf of humans [Boissier 2013]. 
	\paragraph{Agent :} is a computer system situated in some environment, capable of flexible autonomous actions to meet its design objectives [Jennings 1998].
	\\MAS issues have been generally studied in number of specific dimensions [Boissier 2013]
	(Agent, Environment, Interaction, Organisation).\\
	
	\paragraph{Interoperability}
	is a feature of the product or system, whose interfaces are fully understood, to work with other products or systems, now or in the future, both in execution and in access, without any restrictions.
	Although the term is initially defined as information technology or technical systems for information exchange, the broader definition takes into account the social, political and organizational factors that affect system performance. The task of creating coherent services for users when individual components are technically different and managed by different organizations.
	\paragraph{Multi-Agent Interoperability}
	enables heterogeneous agents to communicate among various systems and achieve certain goals on a dynamic basis.
	
	The most important advances in multi-agent engineering systems are the recognition of the importance of the agent environment in which agents are located, through which they interact, as a first-class abstraction. However, existing multi-agent systems based on the environment depend on a predefined static definition of the employee's environment and only those factors that meet this definition can use them. The next powerful step is the idea of an open agent environment that adapts to the response to the agents that occur in it, As in human societies, agent interaction must deal with problems inherent in an open environment: trust and responsibility, dealing with exceptions and social change.
	In theory, these open environments have an open acceptance policy and therefore the potential liquidity membership.\cite{dignum2007open}	
	\paragraph{Openness of MAS Environment}
	This concept is still \underline{open} in literature , i.e. not well defined, but in general it is defined as 	
	\begin{quote}
		``\textit{	the ability of introducing additional
			agents into the system in excess to the agents that comprise it initially. }''\cite{shehory2000software}
	\end{quote} 
	Openness is divided into three levels:
	\begin{enumerate}
		\item Off-line : for example by stopping the system, adding agents, updating some contact information, restarting the system;
		\item Static:  when agents can be added to the system without restart, but both agents notify such an add-on, or they retain an already existing additional potential list
		\item Dynamic: allows agents to leave the system or enter dynamically at runtime without explicit notice.
	\end{enumerate} 
	\paragraph{Sociability in MAS}
	Considering dynamic openness, allowing the system to dynamically adapt to uncertainty in the environment, tasks and availability of sources. As many researchers have pointed out, this kind of uncertainty is especially important for systems that are used in environments with a high degree of dynamism and change, which are currently the rule rather than the exception.
	
	Thus we can refer Environment as a dynamic set of components that we call artifacts , represent functionalities and resources that agents can use to achieve their objectives, and they should be accessible and understandable by agents and other artifacts so an artifact must provide a kind of machine readable manual  \cite{ciortea2016Weaving}
	\paragraph{Social reasoning}
	is defined as any reasoning
	mechanism that uses information about other agents to achieve some inferences \cite{ciortea2016Weaving}\cite{ciortea2016hypermedia} based on Social dependence or social power.
	
	While social dependence mechanism considers the fact that an agent needs to acquire  information about other agents' goals, resources, actions or plans dynamically, The social power mechanism enables reasoning of gains or constraints when entering a group, and both mechanisms can be seen as two sides of the same coin.  \cite{ciortea2016Weaving}
	\paragraph{Web of Things (WoT) :}Approaches, Software Architectural Style, and programming patterns, that allow real world objects to be part of World Wild Web, and in the same way that traditional web provides an application layer to Internet connection layer, WoT acts as an application layer that simplifies creation of IoT applications [Guinard et Al.  2010]. In this context, a Thing can be considered in broad sense of physical objects (sensors, actuators, machines, vehicles,…) and also virtual objects and services

	\paragraph{Semantic Web:} An extension of the World Wide Web through standards by the World Wide Web Consortium (W3C) via Common framework allows data to be shared and reused across application, enterprise, and community boundaries, It can be considered as Integrator across different content, information applications and systems. 
	The term was coined by Tim Berners-Lee for a web of data
	\paragraph{Ontologies}
	Ontology is the representation of common concepts that illustrate the features of being readable automatically, it can be understood and re-use.\cite{chungoora2013model} In addition, it can be used to improve data sharing and reuse, creating a collaborative environments.
	
	Thus, by being familiar with existing ontologies leads to benefit from already represent knowledge to make agents more elegant and compact.
\newpage	\section{Usage Ontology}
	\label{onto}
		\subsection*{Classes}
		
		\subsubsection*{Artifact}
		
		Artifact describes an artifact type where all devices in the environment should be instances of this class or its subclasses
		
		\subsubsection*{Context}
		describes a context or state of an environment entity
		\subsubsection*{Opration}
		operation type where all artifacts operations/actions are instances of this class or its subclasses
		
		\subsubsection*{Usage}
		a usage protocol of operation type of an artifact type specifying its pre/post conditions
		
		\subsection*{Object properties}
		\subsubsection*{forArtifact}
		
		\ensuremath{\exists}~forArtifact~Thing~\ensuremath{\sqsubseteq}~Usage
		
		\ensuremath{\top}~\ensuremath{\sqsubseteq}~\ensuremath{\forall}~forArtifact~Artifact
		
		this usage is valid only for operations related to specific artifact
		\subsubsection*{forOperation}
		
		\ensuremath{\exists}~forOperation~Thing~\ensuremath{\sqsubseteq}~Usage
		
		\ensuremath{\top}~\ensuremath{\sqsubseteq}~\ensuremath{\forall}~forOperation~Opration
		
		the usage is valid for specific operation
		\subsubsection*{hasContext}
		
		\ensuremath{\exists}~hasContext~Thing
		
		\ensuremath{\top}~\ensuremath{\sqsubseteq}~\ensuremath{\forall}~hasContext~Context
		
		entity environment has a context
		\subsubsection*{hasOperation}
		
		\ensuremath{\exists}~hasOperation~Thing~\ensuremath{\sqsubseteq}~Artifact
		
		\ensuremath{\top}~\ensuremath{\sqsubseteq}~\ensuremath{\forall}~hasOperation~Opration
		
		the artifact has an operation
		\subsubsection*{hasPostcond}
		
		\ensuremath{\exists}~hasPostcond~Thing~\ensuremath{\sqsubseteq}~Usage
		
		\ensuremath{\top}~\ensuremath{\sqsubseteq}~\ensuremath{\forall}~hasPostcond~Context

		effects of usage protocol on environment by meens of a new context may be achieved after calling this usage operation
		\subsubsection*{hasPrecond}
		
		\ensuremath{\exists}~hasPrecond~Thing~\ensuremath{\sqsubseteq}~Usage
		
		\ensuremath{\top}~\ensuremath{\sqsubseteq}~\ensuremath{\forall}~hasPrecond~Context
		
		a context  that should present in the envirnment to call the operation in this usage protocol
		\subsection*{Data properties}\subsection*{Individuals}\subsection*{Datatypes}\subsubsection*{PlainLiteral}
	\newpage
	\section{SPARQL Update Queries}
	 SPARQL 1.1 Update is an update language for RDF graphs.\cite{SparqlW3C}
	\\The Graph update operations modify the existing Graphs in the Graph Store, but do not delete them explicitly and do not create them.\cite{SparqlW3C}
	\\However, nonempty insertions in non-existent graphs implicitly create these graphs, that is, an implementation that satisfies an update request SHOULD silently create graphs that do not exist before the insertion of triplets. and MUST return with a failure if it fails to do so for any reason.\\ (For example, the implementation may have insufficient resources, or an implementation may only provide an update service on a fixed set of graphs and the graph implicitly created is not in this fixed set).\\ An implementation MAY delete the graphs left empty after the deletion of the triplets.
	\\The DELETE / INSERT operation can be used to remove or add triplets from/to the graph store, based on bindings for a query template specified in a WHERE clause, that will be used to calculate the sequence of the solution from the bindings to a set of variables.\\ The links for each solution are then replaced in the DELETE template to remove the triple, and then in the INSERT template to create a new triple.\\ If any solution produces a triple that contains an unbound variable or an illegal RDF construct, such as a literal in a subject or predicate position, that triple is not included when calculating the operation: INSERT will not create new data in the resulting graph, and DELETE will not remove Anything. The graphs used to calculate the solution sequence may differ from the modified graphs with the DELETE and INSERT templates.
		
	\section{Implementation Class Diagrams}
	\label{classes}
	\begin{figure}[ht]
		\centering
		\includegraphics[width=\linewidth]{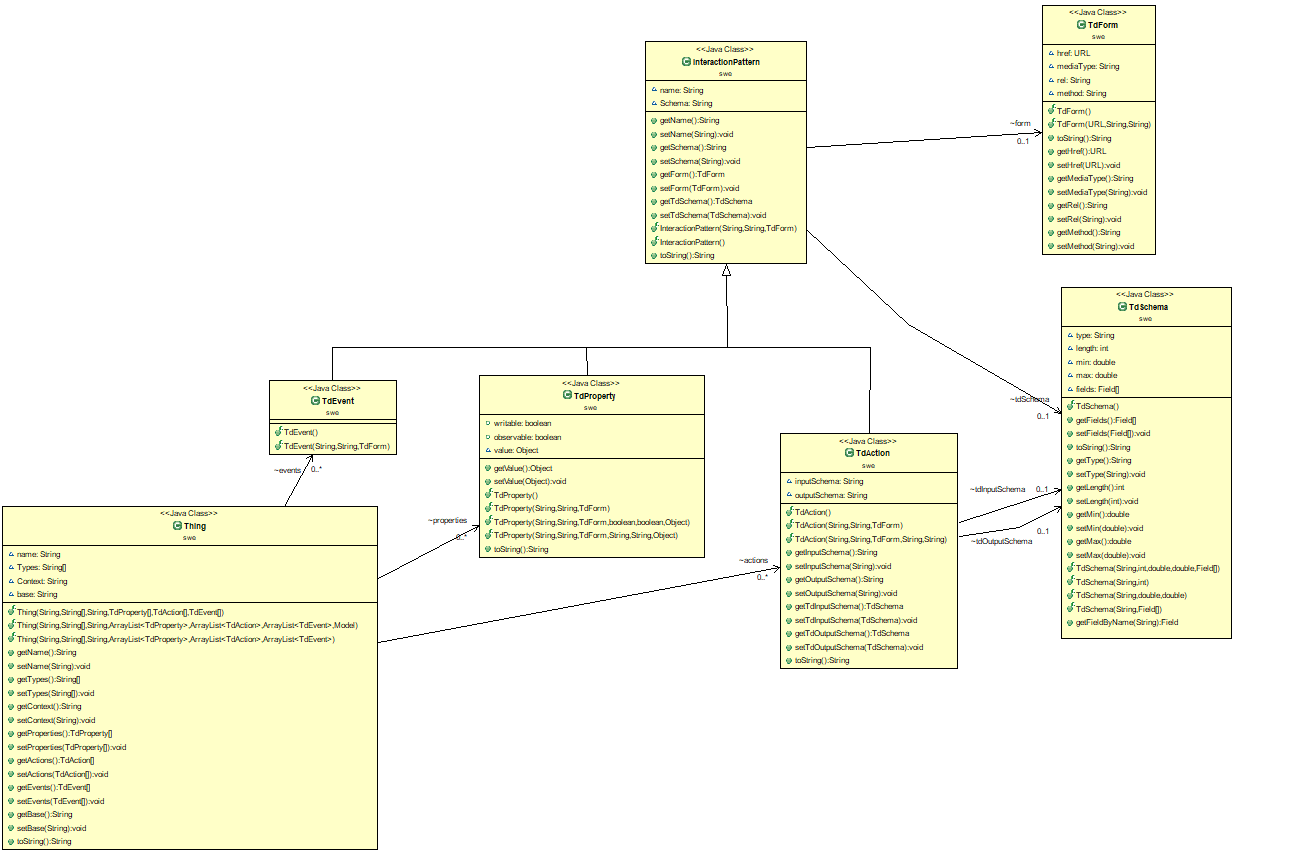}
		\caption{Thing hierarchy}
		\label{fig:thingcd}
	\end{figure}
	\begin{figure}[ht]
		\centering
		\includegraphics[width=\linewidth]{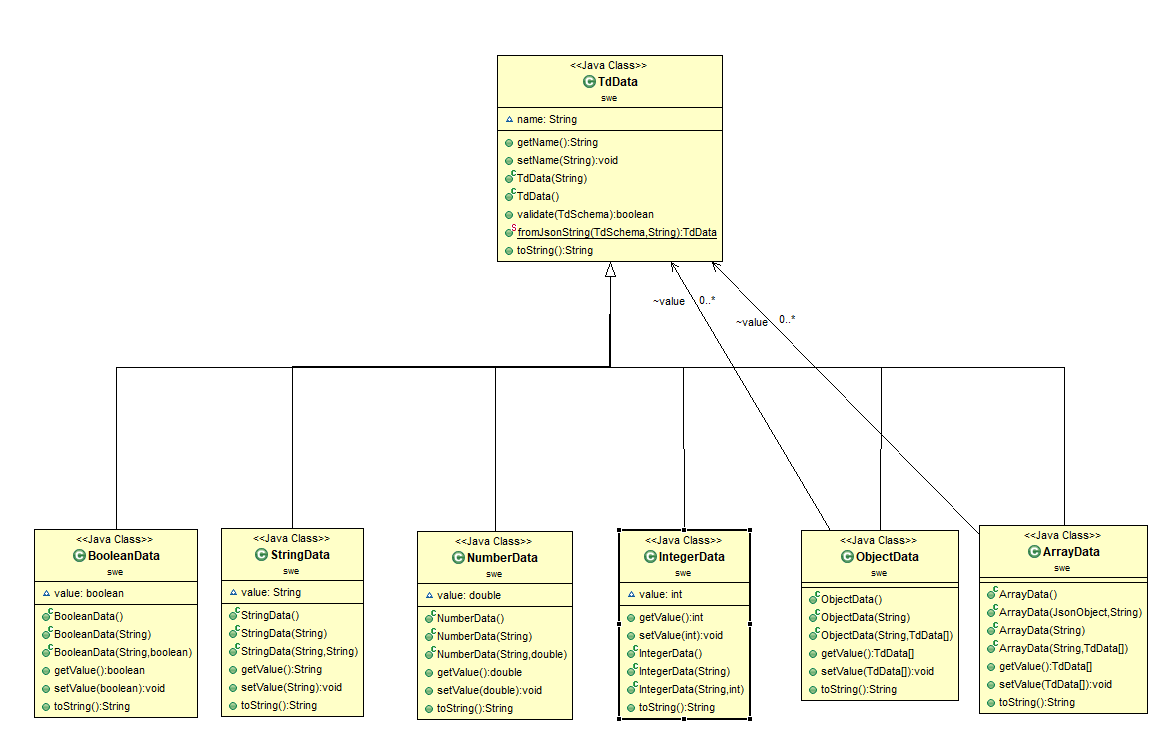}
		\caption{Data hierarchy}
		\label{fig:data}
	\end{figure}

\end{document}